\documentclass[]{spie}  %>>> use for US letter paper
\usepackage{amsmath,amsfonts,amssymb}
\usepackage{graphicx}
\usepackage{multirow}
\usepackage[colorlinks=true, allcolors=blue]{hyperref}

\title{The Geant4 mass model of the ATHENA Silicon Pore Optics and its effect on soft proton scattering}

\author[a]{Valentina Fioretti}
\author[a]{Andrea Bulgarelli}
\author[b]{Simone Lotti}
\author[b]{Claudio Macculi}
\author[c]{Teresa Mineo}
\author[b]{Luigi Piro}
\author[a]{Debora Bruno}
\author[a]{Massimo Cappi}
\author[a]{Mauro Dadina}

\affil[a]{INAF Osservatorio di Astrofisica e Scienza dello spazio, Via P. Gobetti 101, 40129, Bologna, Italy}
\affil[b]{INAF Istituto di Astrofisica e Planetologia Spaziali, Via Fosso del Cavaliere, 100, Roma, Italy}
\affil[c]{INAF Istituto di Astrofisica Spaziale e Fisica Cosmica di Palermo, Via U. La Malfa 153, 90146 Palermo, Italy}

\authorinfo{Further author information: (Send correspondence to V. Fioretti)\\E-mail: fioretti@iasfbo.inaf.it, Telephone: +39 051 6398772}

\begin{document} 
\maketitle

\begin{abstract}
Given the unprecedented effective area, the new ATHENA Silicon Pore Optics (SPO) focusing technology, the dynamic and variable L2 environment, where no X-ray mission has flown up to date, a dedicated Geant4 simulation campaign is needed to evaluate the impact of low energy protons scattering on the ATHENA mirror surface and the induced residual background level on its X-ray detectors. 
The Geant4 mass model of ATHENA SPO is built as part of the ESA AREMBES project activities using the BoGEMMS framework. An SPO mirror module row is the atomic unit of the mass model, allowing the simulation of the full structure by means of 20 independent runs, one for each row. 
No reflecting coating is applied in the model: this simplification implies small differences (few percentages) in the proton flux, while reducing the number of volumes composing the mass model and the consequent simulation processing time. 
Thanks to the BoGEMMS configuration files, both single pores, mirror modules or the entire SPO row can be built with the same Geant4 geometry. 
The conical approximation used for the Si plates transmits 20\% less photons than the actual SPO design, simulated with a ray-tracing code. Assuming the same transmission reduction for protons, a 20\% uncertainty can be accepted given the overall uncertainties of the input fluxes.
Both Remizovich, in its elastic approximation, and Coulomb single scattering Geant4 models are used in the interaction of mono-energetic proton beams with a single SPO pore. The scattering efficiency for the first model is almost twice the efficiency obtained with the latter but for both cases we obtain similar polar and azimuthal angular distributions, with about 70-75\% of scatterings generated by single or double reflections. The soft proton flux modelled for the plasma sheet region is used as input for the simulation of soft proton funnelling by the full SPO mass model. A much weaker soft proton vignetting than the one observed by XMM-Newton EPIC detectors is generated by ATHENA mirrors. The residual soft proton flux reaching the focal plane, defined as a 15 cm radius, is $10^{4}$ times lower than the input L2 soft proton population entering the mirror, at the same energy, with rates comparable or higher than the ones observed in XMM EPIC-pn most intense soft proton flares.

\end{abstract}


\section{Introduction}
The ATHENA space telescope is the future ESA L-class X-ray mission, designed to address the Cosmic Vision \textit{The Hot and Energetic Universe} science theme.  With a planned launch on 2030 in L2 orbit, ATHENA will carry, as X-ray telescope, a modular mirror based on Silicon Pore Optics (SPO) technology with a focal length of 12 m and an external diameter, for the baseline design, up to 3 m. Two instruments populate the focal plane, covering the soft $<15$ keV energy range: a Wide Field Imager (WFI\cite{2017SPIE10397E..0VM}) for wide field imaging and spectroscopy and an X-ray Integral Field Unit (X-IFU\cite{2016SPIE.9905E..2GL}) for fine X-ray spectroscopy.
\\
For X-ray telescopes operating outside the radiation belts, low energy protons ($< 300$ keV) can enter the field of view, scatter with the Wolter-I reflecting surface, and be funneled to the detection plane. The issue of soft proton scattering by X-ray optics has been known since the launch of Chandra and XMM-Newton missions, with induced loss of charge transfer efficiency on Chandra ACIS front-illuminated CCDs\cite{2000SPIE.4140...99O} and the detection of intense background flares by the XMM EPIC\cite{2007A&A...464.1155C, 2008A&A...486..359L} instruments. 
Given the large ATHENA effective area and the new SPO focusing technology, dedicated Geant4 simulations are mandatory to evaluate the impact of low energy proton funneling through the mirrors on the X-IFU and WFI sensitivity and, if needed, driving the design of a magnetic diverter. 
Geant4 simulations of the XMM EPIC-pn soft proton induced background produced in the past\cite{2016SPIE.9905E..6WF} underestimated the observations because of the large variability of the flares and the need for long term monitoring of such events to produce a mission averaged X-ray soft proton spectrum. 
\\
The SPO technology	\cite{2017SPIE10399E..0CC} is based on Silicon (Si) wafers cut by parallel grooves to create ribs. The stacking of the ribbed Si plates, coated by reflective metals, generates millions of pores, through which the X-rays, and potentially charged particles, propagate. One mirror module comprises two parabolic and hyperbolic stacks, mounted in a common bracket, to create the Wolter-I reflecting configuration.
\\
Building the Geant4 SPO mass model, with hundreds of millions of volumes, and achieving a simulation of the whole structure interacting with the soft proton population while keeping a feasible CPU processing time, represents a challenge in itself. The ESA AREMBES (ATHENA Radiation Environment Models and X-ray Background Effects Simulators) project\footnote{http://space-env.esa.int/index.php/news-reader/items/AREMBES.html} will deliver to the science community a Geant4-based framework for the simulation of ATHENA X-ray background and shielding efficiency, including the SPO mass model presented here (see Sec. \ref{sec:mass_model}). The verification of the SPO implementation is an opportunity to compare the effect of the currently proposed scattering models, single scattering and the elastic Remizovich solution, on the distribution of protons exiting the optics (Sec. \ref{sec:ver}).
We rely here on the L2 soft proton environment modeling\cite{lotti_2018_submitted} achieved in the ESA AREMBES project to estimate the input proton population entering the SPO pores and study the residual proton flux expected on the focal plane (Sec. \ref{sec:full}).

\section{SPO mass model}\label{sec:mass_model}
\begin{figure} [h!]
   \begin{center}
   \begin{tabular}{c} %% tabular useful for creating an array of images 
   \includegraphics[width=0.4\textwidth]{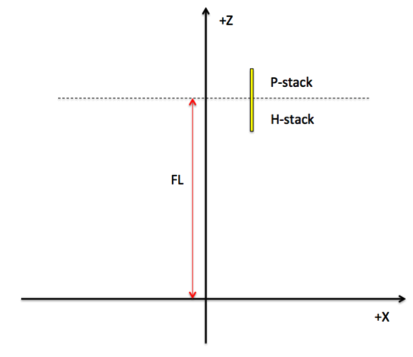}
   \includegraphics[width=0.5\textwidth]{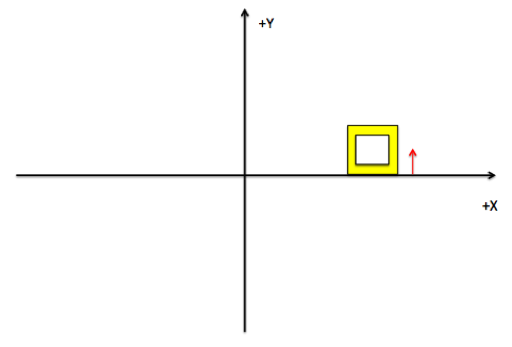}
   \end{tabular}
   \end{center}
   \caption 
%>>>> use \label inside caption to get Fig. number with \ref{}
   { \label{fig:place}SPO stack placement in the Geant4 reference system (left panel) and single pore placement in the X-Y plane (right panel).}
   \end{figure}

The SPO simulation described here refers to the baseline configuration\cite{2015SPOreport} , composed by 20 rows. In this configuration the pore radii are calculated using a conical approximation, but their differences from the true Wolter-I values are below the $\mu\rm m$ level.
Thanks to the modular design of the Geant4 mass model, any row can be removed from the mass model.
\\
One mirror module (MM) comprises two Paraboloid-Hyperboloid (P-H) stacks, mounted in a common bracket. %Each stack has 35 Si plates, of which 34 are reflecting, for a total of 68 reflecting plates in each mirror module. 
The MM circumferential width is variable, as the total number of MMs composing a single row. 
The following simplifications are applied to the SPO design:
\begin{itemize}
\item no distance is placed between two stacks composing an MM;
\item all mirror modules are placed on the same plane, normal to the telescope axis, contrary to the real placement on a spherical plane, which moves the modules up to $\sim10$ cm closer to the focal plane to improve the off-axis PSF performance; 
\item no coating is applied to the Si plates (see Sec. \ref{sec:coating} for details);
\item the distance between the plates is constant and uses the distance between the first and the second reflective plates\cite{2015SPOreport} .
\end{itemize}
The optics symmetry axis is the Z-axis of the Geant4 reference system and the focal plane is placed on the X-Y plane at Z = 0 (Fig. \ref{fig:place}, left panel). The focal length starts at the intersection between the paraboloid and the hyperboloid Wolter-I sections. The position of the first pore of the first mirror module with respect to the X-Y plane is shown in Fig. \ref{fig:place} (right panel).

\subsection{Geant4 implementation}
 \begin{figure} [h!]
   \begin{center}
   \begin{tabular}{c} %% tabular useful for creating an array of images 
   \includegraphics[width=0.45\textwidth]{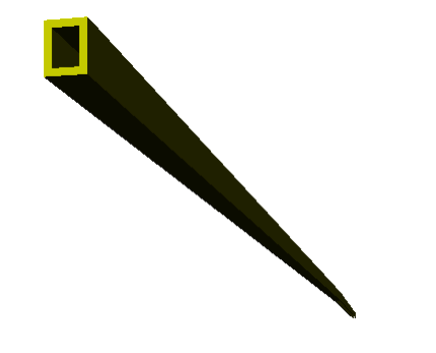}
   \includegraphics[width=0.45\textwidth]{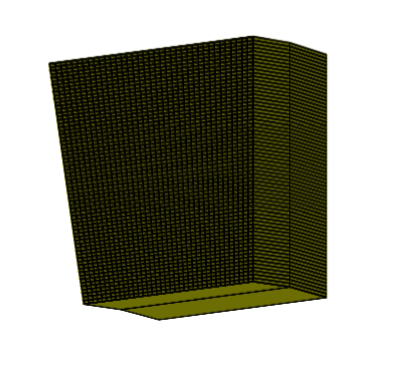}\\   
   \includegraphics[width=0.45\textwidth]{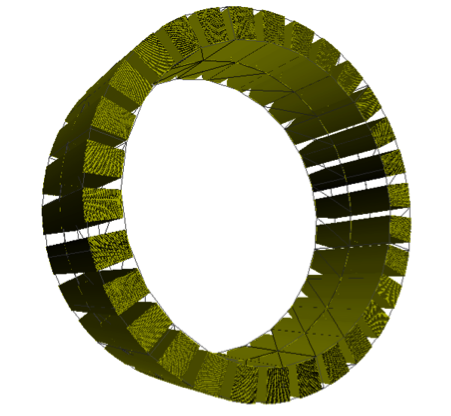}
    \includegraphics[width=0.38\textwidth]{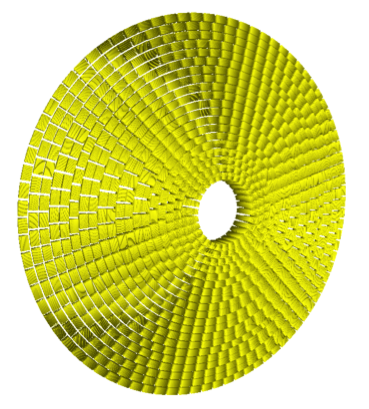}\\
   \end{tabular}
   \end{center}
   \caption 
%>>>> use \label inside caption to get Fig. number with \ref{}
   { \label{fig:model}\textit{Top panels:} Geant4 model for one pore (left panel) and a full MM (right panel). \textit{Bottom panels:} Geant4 model for a single row (left panel) and the full SPO model obtained by merging the 20 independent rows.}
   \end{figure} 
The Geant4 \cite{g4_1, g4_2, 2016NIMPA.835..186A} toolkit, initially developed by CERN and then mantained by a large collaboration, is a C++ based particle transport code for the simulation of high energy experiments at particle accelerators and then extended to lower energies (sub-keV scale). BoGEMMS (Bologna Geant4 Multi-Mission simulator) is a Geant4-based simulation framework \cite{2012SPIE.8453E..35B} for the evaluation of the scientific performance (e.g. background spectra, effective area) of high energy experiments, with particular focus on X-ray and gamma-ray space telescopes. The BoGEMMS framework is used throughout the activity to build the Geant4 geometry and convert it to a GDML\footnote{Geometry Description Markup Language, a specialized XML-based language designed for the description of Geant4 volumes.} format as part of the AREMBES ATHENA mass model. The BoGEMMS ability to customize the geometry from a configuration file is used here to set the SPO design at run-time. This feature will easily allow to modify, if needed, the AREMBES SPO mass model during the validation phase.
\\
All volumes composing the MMs, both plates and ribs, are built using the Geant4 parameterization class using a cone segment as base volume. The mother volume containing the parameterized plates and ribs is a ring including both the paraboloid and the hyperboloid sections.
BoGEMMS configuration file allows to build at run-time, using the same Geant4 geometry, a single pore, a single MM or an entire SPO row (Fig. \ref{fig:model}). The MM row is the atomic unit of the AREMBES ATHENA SPO mass model. It allows to run 20 independent simulations, one for each SPO row, of the charged particle interaction with the optics surface, minimizing the CPU processing time while simulating the full structure. If rows are removed because of downgrades of the ATHENA mission, AREMBES mass model is still valid. The full SPO structure is shown in Fig. \ref{fig:model} (bottom-right panel), obtained by merging the 3D models of the 20 rows.
% \begin{figure} [h!]
  % \begin{center}
 %  \begin{tabular}{c} %% tabular useful for creating an array of images 
 %  \includegraphics[width=0.3\textwidth]{./plots/fig4_left.png}
 %  \includegraphics[width=0.25\textwidth]{./plots/fig4_center.png}
 %  \includegraphics[width=0.05\textwidth]{./plots/fig4_right.png}
 %  \end{tabular}
 %  \end{center}
 %  \caption 
%>>>> use \label inside caption to get Fig. number with \ref{}
%   { \label{fig:maya}Full Geant4 SPO model if all the rows are built in the same simulation by merging the 20 independent row models.}
 %  \end{figure} 

\section{Soft protons scattering on ATHENA SPO}\label{sec:ver}
The current Geant4 implementation of the ATHENA SPO mass model has been tested by simulating the interaction of low energy protons with a simple slab, a single pore, and the full SPO structure. 
The interaction of soft protons with the SPO volumes has been simulated with the 10.01.p02 and 10.02.p03 releases but no significant differences were found (all results here refer to the 10.2 release).
From the physics validation activity\cite{2017ExA....44..413F} , where Geant4 simulations where compared to laboratory measurements of low energy proton scatterings by a sample of e-Rosita X-ray mirror shells, two physics models are used: 
\begin{itemize}
\item the Remizovich solution\cite{1983RadEffect, 1980JETP...52..225R} describes particles reflected by solids at glancing angles in terms of the Boltzmann transport equation using the diffuse approximation and the model of continuous slowing down in energy. The Geant4 implementation\cite{2017ExA....44..413F} , one of the products of AREMBES activities, uses the approximated formula in the assumption of no energy losses.
\item In the Coulomb, or Rutherford, scattering the particle, traversing the medium, undergoes one or more elastic scatterings with the electron field of the nuclei. For a grazing incident angle, as the case of low energy protons funnelled by X-ray optics, the particle can escape the surface after few interactions. The use of the single scattering (SS) model, as provided by the Geant4 reference \textit{G4EmStandardPhysicsSS} physics list, allows the computation of each single scattering, contrary to the multiple scattering model that averages the proton energy and angle over a larger number of interactions. Throughout the text, the Geant4 reference \textit{G4EmStandardPhysicsSS} physics list is used when not specified.
\end{itemize}
Since the new physics list produced by the AREMBES collaboration, optimized for X-ray space applications, uses the Coulomb single scattering for proton energies below 1 MeV, the results obtained with the SS physics list can also represent a verification test for the AREMBES \textit{SPACE} physics list optimized for space applications. The production cut applied to all simulations is 1 $\mu m$. 

\subsection{Effect of coating}\label{sec:coating}
\begin{figure} [h!]
   \begin{center}
   \begin{tabular}{c} %% tabular useful for creating an array of images 
   \includegraphics[width=0.49\textwidth]{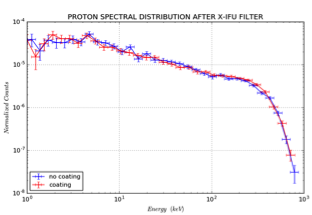}
   \includegraphics[width=0.49\textwidth]{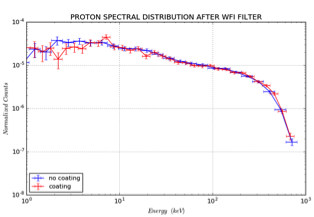}
   \end{tabular}
   \end{center}
   \caption 
%>>>> use \label inside caption to get Fig. number with \ref{}
   { \label{fig:coating}Proton spectral distribution after scattering with the slab and crossing the X-IFU (left panel) and the WFI (right panel) optical filters, with and without coating.}
   \end{figure}

The reflective side of the Si plates is coated by a composite layer of 10 nm of Iridium and 8 nm of B4C. Adding the coating volumes to the SPO mass model increases the complexity and the overall CPU processing time.
The impact of the coating on the proton distribution is evaluated by simulating the interaction with a planar Si slab, with the same thickness of SPO Si plates, with and without the coating layers. Since the Remizovich solution, in the approximation of no energy losses, does not depend on the reflecting material, hence the result does not change with or without the coating, the test is only run with the SS model. 
\\
A power-law spectrum in the 10 keV -- 1 MeV energy range is used as input flux.
The proton energy distribution after the scattering by the slab and the proton crossing through the optical filters covering the detectors is shown in Fig. \ref{fig:coating}, with the red and blue crosses referring to the coating and no-coating configuration respectively. 
The effect is shown for the use of the X-IFU\cite{2017ExA....44..371L} (left panel) and WFI\cite{2015SPIE.9601E..09B} (right panel) optical filters: the difference in the number of proton surviving the interaction with the optical filter, with or without coating, is of few percents.  It is possible to remove the coating from the Geant4 mass model, lowering the CPU processing time, without significantly losing accuracy in the simulated proton distribution at the ATHENA focal plane\cite{2017BRUNOreport}. 

\subsection{Single pore simulations}
The first pore of the inner SPO row is built as a stand-alone geometry to test the correct interaction of protons with the Si pores composing the mirror modules. Protons are simulated from a distance of 10 $\mu m$ with respect to the pore entrance, from a planar surface equal to the pore entrance, within a 1$^{\circ}$ semi-aperture cone and using a cosine-law angular distribution. Three mono-energetic beams (100, 250 and 500 keV) are used to map a potential dependency of the transmission on the input proton energy (only the 100 keV case is shown for the Remizovich model since it does not depend on the proton energy).
 \begin{figure} [h!]
   \begin{center}
   \begin{tabular}{c} %% tabular useful for creating an array of images 
   \includegraphics[width=0.7\textwidth]{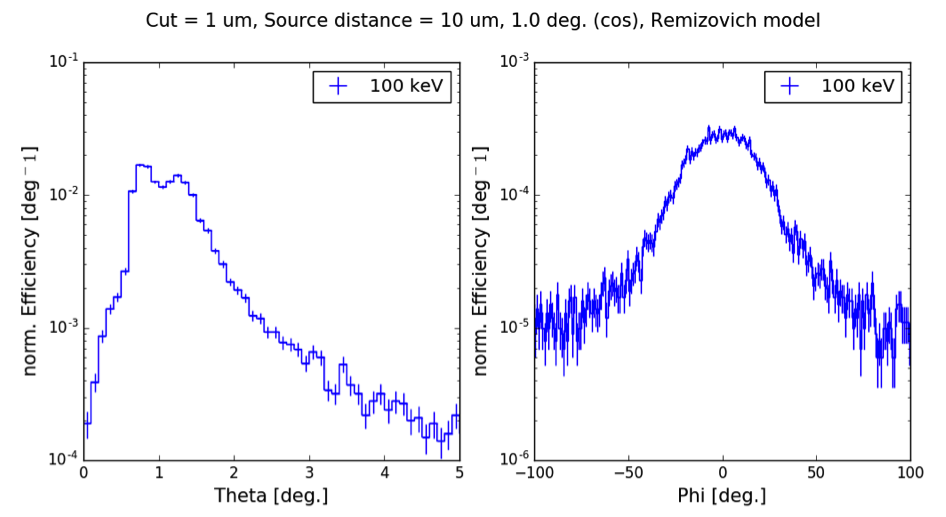}
   \end{tabular}
   \end{center}
   \caption 
%>>>> use \label inside caption to get Fig. number with \ref{}
   { \label{fig:rem}Polar (left panel) and azimuthal (right panel) scattering efficiency distribution, normalized for the bin width, if the Remizovich model is used.}
   \end{figure} 
 \begin{figure} [h!]
   \begin{center}
   \begin{tabular}{c} %% tabular useful for creating an array of images 
   \includegraphics[width=0.7\textwidth]{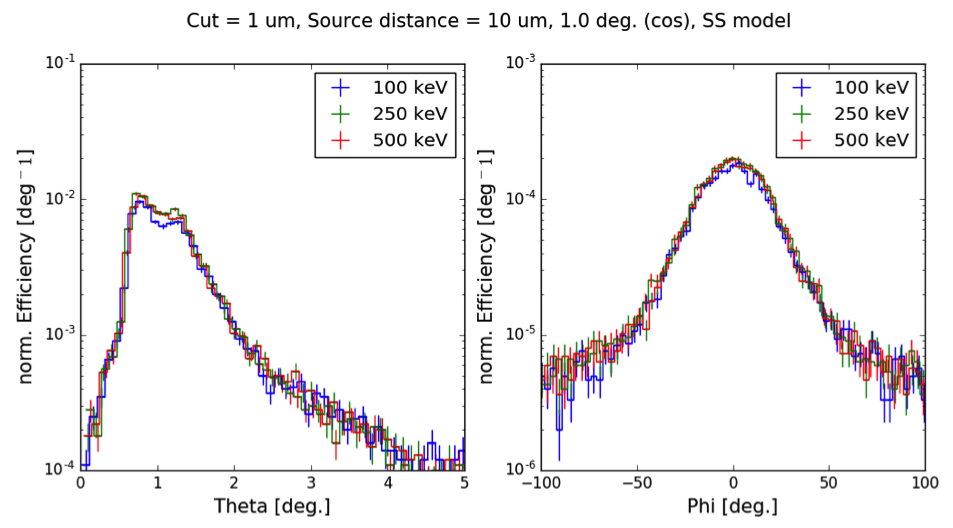}
   \end{tabular}
   \end{center}
   \caption 
%>>>> use \label inside caption to get Fig. number with \ref{}
   { \label{fig:ss}Polar (left panel) and azimuthal (right panel) scattering efficiency distribution, normalized for the bin width, if the SS model is used.}
   \end{figure} 
If we define the scattering efficiency as the fraction of exiting to entering protons, we obtain for mono-energetic protons beams (statistical errors are omitted being much smaller then the values):
\begin{itemize}
\item Remizovich (no energy losses) model: 0.017
\item SS model:
\subitem - 100 keV: 0.0090
\subitem - 250 keV: 0.0097
\subitem - 500 keV: 0.0101
\end{itemize}
As expected from the physics validation study\cite{2017ExA....44..413F} , the Remizovich model - in the approximation of no energy losses - results in almost a factor 2 higher number of scattered protons. A $\sim10\%$ increase in the scattering efficiency is observed if we increase the proton energy using the SS model. The scattering efficiency, normalized for the bin width, as a function of the polar (\textit{Theta}) and azimuthal (\textit{Phi}) angle of the protons exiting the pore is shown in Fig. \ref{fig:rem} (Remizovich model) and Fig. \ref{fig:ss} (SS model). If the polar angle is 0, the proton travels along the telescope axis. 
\\
A major result of the present verification test is that the same angular distribution for the protons exiting the pore, although with different integral values, is obtained for the two physics models under study. While the azimuthal angle would be uniformly distributed in the $\left[ -180^{\circ}-180^{\circ} \right]$ range if the full SPO is used, simulating a single pore results into the azimuthal angle peaking towards the direction where the pore is inclined (Fig. \ref{fig:rem} and \ref{fig:ss}, right panel). Given that the plate inclination with respect to the telescope axis is 0.3$^{\circ}$ or 0.9$^{\circ}$, depending on the reflecting paraboloid or hyperboloid surface respectively, the maximum specular scattering angle is about 1.6$^{\circ}$ and 2.8$^{\circ}$ respectively. The polar scattering angle distribution peaks below 2$^{\circ}$, confirming the higher scattering probability at specular angles, with the two peaks mainly caused by single or double reflections. 
 \begin{figure} [h!]
   \begin{center}
   \begin{tabular}{c} %% tabular useful for creating an array of images 
   \includegraphics[width=0.65\textwidth]{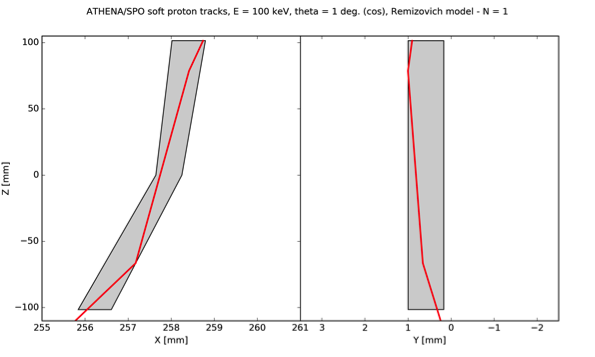}
   \includegraphics[width=0.35\textwidth]{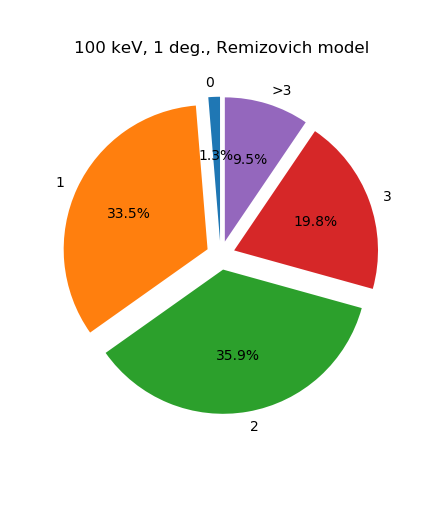}
   \end{tabular}
   \end{center}
   \caption 
%>>>> use \label inside caption to get Fig. number with \ref{}
   { \label{fig:rem_scatt}Visualization of the proton scattering behavior for one single event along the pore (left panel) and distribution of the number of proton reflections, within the same event, if the Remizovich model is used.}
   \end{figure} 
   \begin{figure} [h!]
   \begin{center}
   \begin{tabular}{c} %% tabular useful for creating an array of images 
   \includegraphics[width=0.65\textwidth]{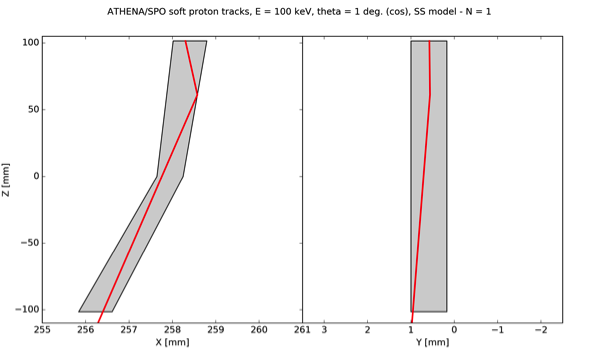}
   \includegraphics[width=0.35\textwidth]{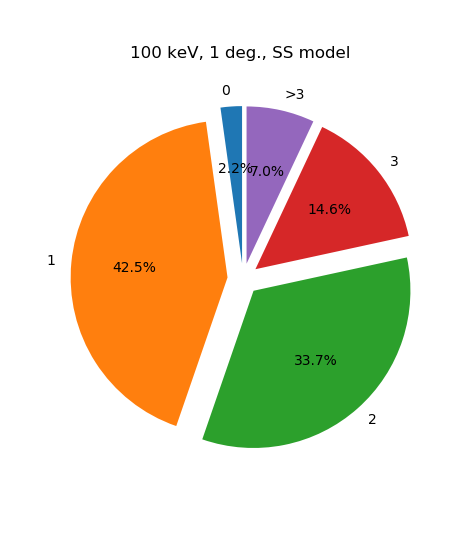}
   \end{tabular}
   \end{center}
   \caption 
%>>>> use \label inside caption to get Fig. number with \ref{}
   { \label{fig:ss_scatt}Visualization of the proton scattering behavior for one single event along the pore (left panel) and distribution of the number of proton reflections, within the same event, if the SS model is used.}
   \end{figure} 
    \begin{figure} [h!]
   \begin{center}
   \begin{tabular}{c} %% tabular useful for creating an array of images 
   \includegraphics[width=0.44\textwidth]{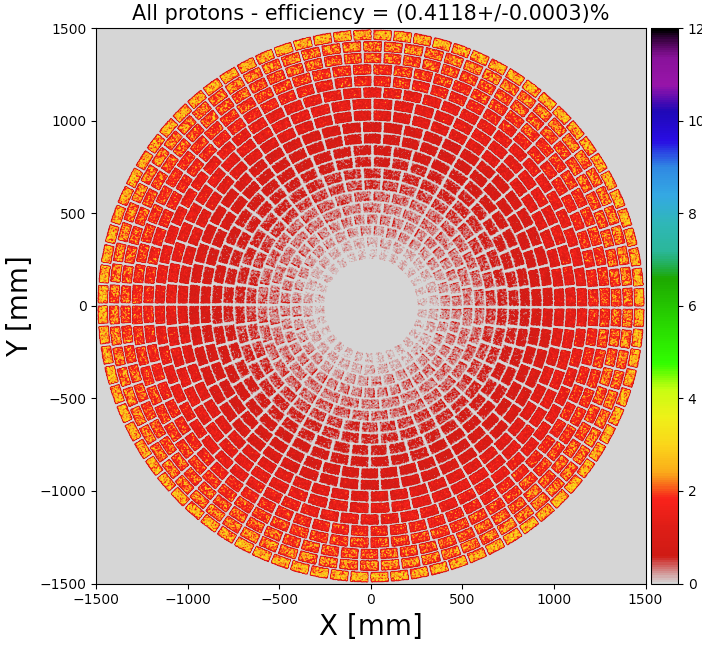}
   \includegraphics[width=0.48\textwidth]{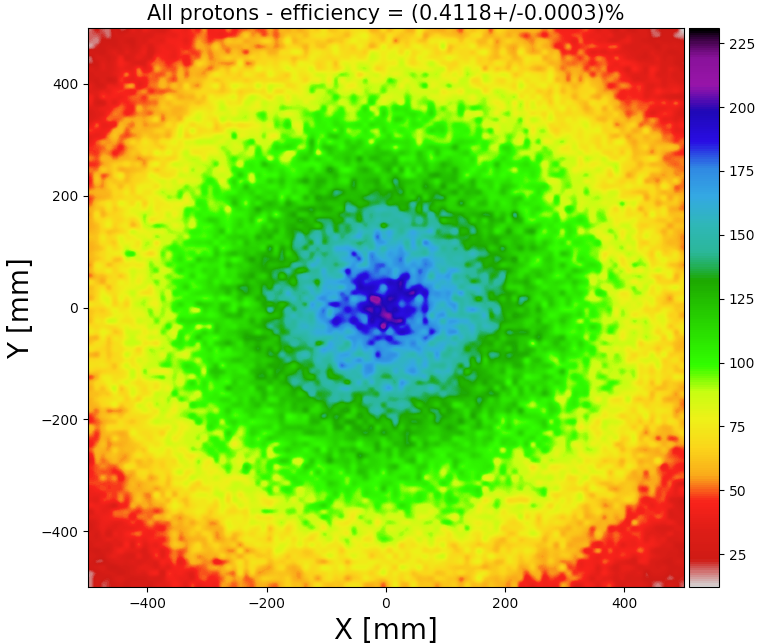}
   \end{tabular}
   \end{center}
   \caption 
%>>>> use \label inside caption to get Fig. number with \ref{}
   { \label{fig:spatial}Spatial distribution at the SPO exit (left panel) and at focal plane (right panel) for all protons exiting the mirror.}
 \end{figure} 
The proton scattering within the pore is visualized in Fig. \ref{fig:rem_scatt} and \ref{fig:ss_scatt} (left panel) from the lateral X and Y views: the grey area refers to the pore inner surface and the red line is the proton track. This simulation quick-look viewer allows to check the correctness of the track analysis and the overall physics interaction in the verification phase.
The distribution of the number of proton reflections (Fig. \ref{fig:rem_scatt} and \ref{fig:ss_scatt}) obtained for the two models shows that in about 70\% of the events we get 1 or 2 reflections, with more single interactions if SS is used. A small percentage of protons ($<2.2\%$) exit the pore without interacting with the Si surface. 

\subsection{Full SPO simulations}\label{sec:full}
\begin{figure} [t]
   \begin{center}
   \begin{tabular}{c} %% tabular useful for creating an array of images 
   \includegraphics[width=0.4\textwidth]{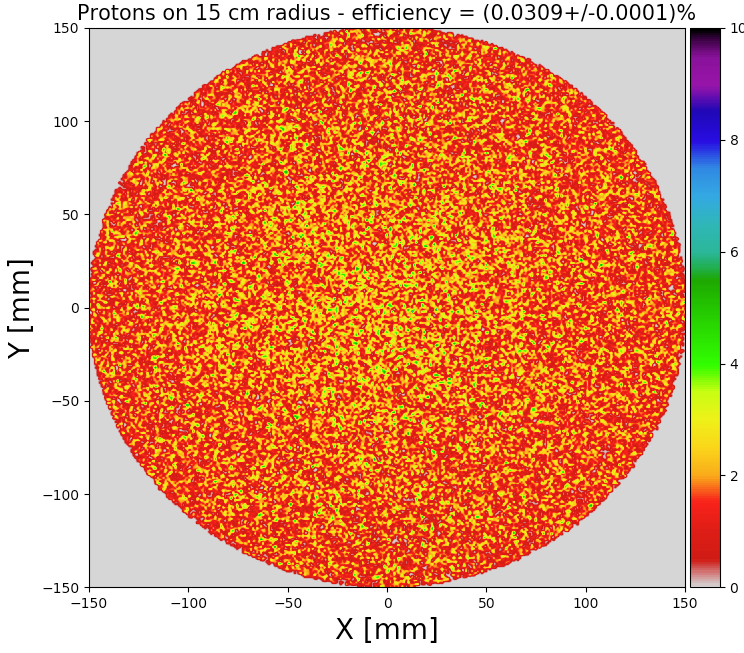}
   \includegraphics[width=0.56\textwidth]{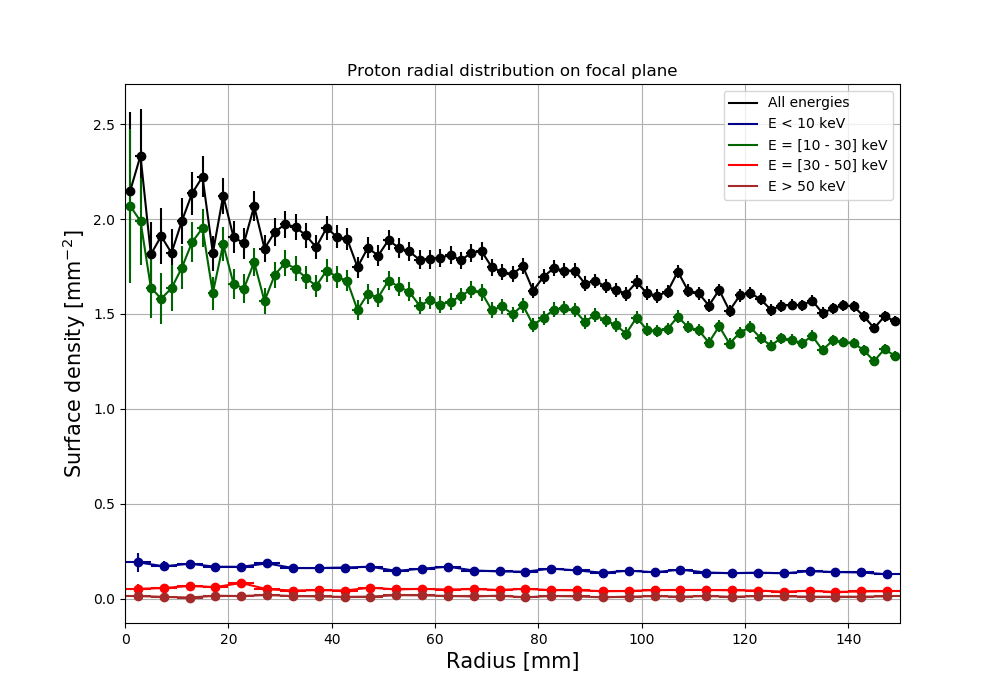}
   \end{tabular}
   \end{center}
   \caption 
%>>>> use \label inside caption to get Fig. number with \ref{}
   { \label{fig:radial}Proton spatial (left panel) and radial (right panel) distribution within a 15 cm radius region on the focal plane.}
 \end{figure}

\begin{figure} [h!]
   \begin{center}
   \begin{tabular}{c} %% tabular useful for creating an array of images 
   \includegraphics[width=0.38\textwidth]{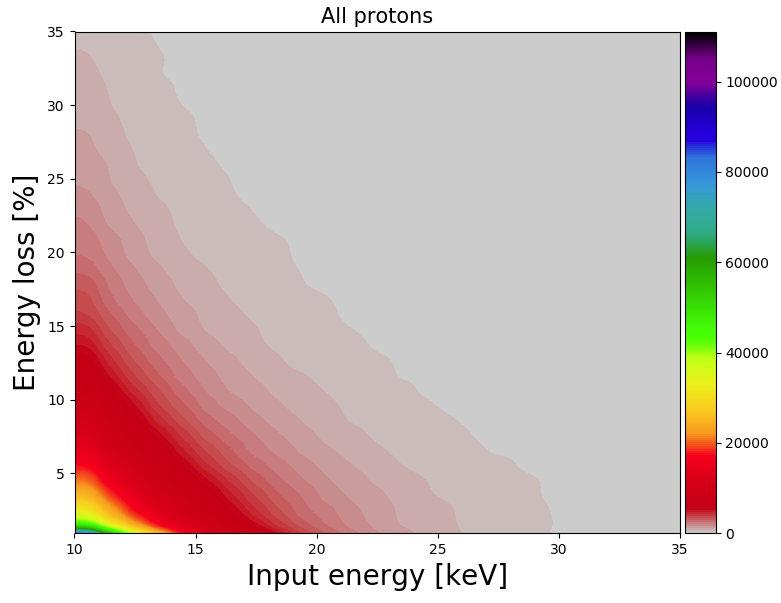}
   \includegraphics[width=0.4\textwidth]{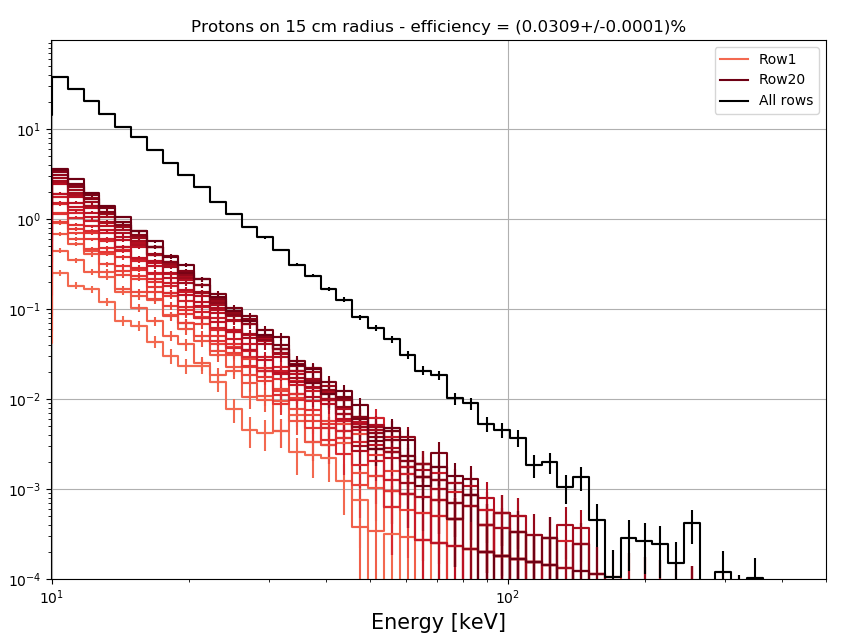}\\
   \includegraphics[width=0.8\textwidth]{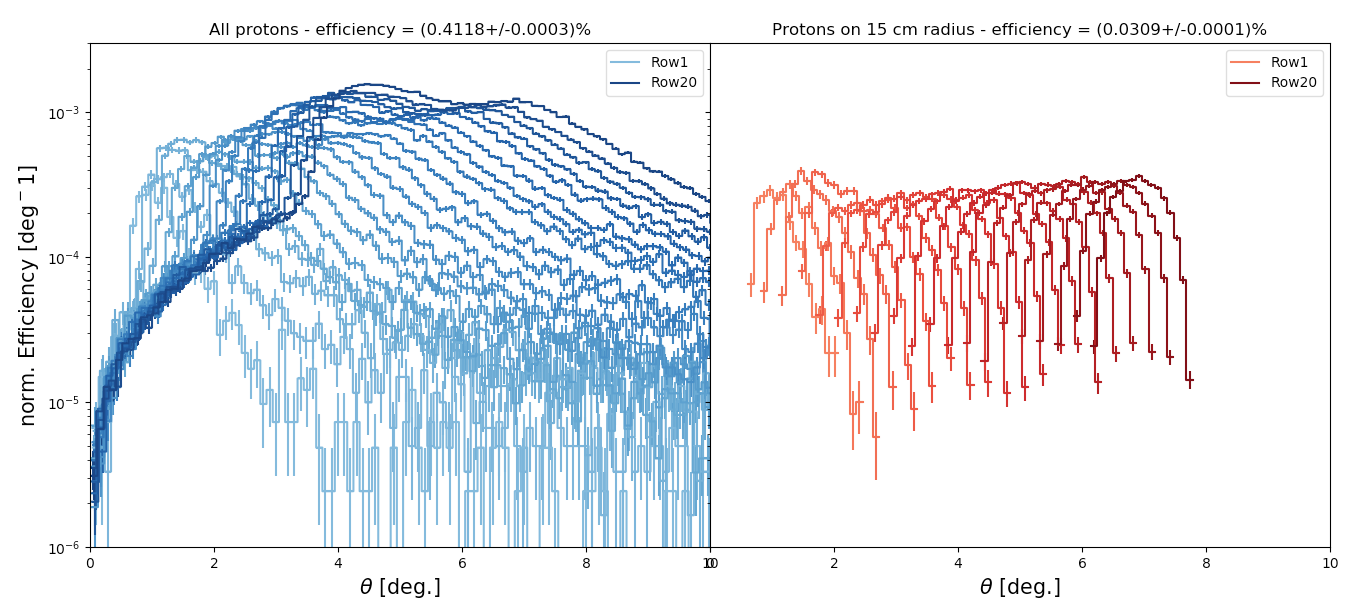}
   \end{tabular}
   \end{center}
   \caption 
%>>>> use \label inside caption to get Fig. number with \ref{}
   { \label{fig:energy}\textit{Top panel:} energy loss, in percentage, as a function of the input energy for all protons exiting the SPO (left panel) and energy distribution of protons reaching the focal plane, normalized for the selected 15  cm radius area (right panel). \textit{Bottom panel:} polar angle distribution for all protons exiting the SPO (left panel) and only the protons reaching the focal plane within a 15 cm radius.}
 \end{figure}

The full SPO implementation uses a representative model of the input proton flux expected in L2 orbit. Following a conservative approach, we chose the magnetotail plasma regime modelled for the plasma sheet, a \textit{sheet-like} region in the equatorial magnetotail characterized by the highest soft proton intensity, resulting from the analysis of the AREMBES project\cite{WP2_report2}.
A power-law model, from 10 keV to 500 keV, describes the energy distribution, with a $\pi$ sr integrated flux of $\sim4\times10^{5}$ prot. cm$^{-2}$ s$^{-1}$ keV$^{-1}$ at 10 keV corresponding to a cumulative fraction of 90\%, i.e. 90\% of the time the SPO are exposed to a flux below this value. The solar wind powered low energy proton flux measured by Equator-S along XMM-Newton orbit, with a value of 3 prot. cm$^{-2}$ s$^{-1}$ keV$^{-1}$ sr$^{-1}$ at 100 keV\cite{2016SPIE.9905E..6WF} , is 5 times lower than the L2 model used as reference in the present simulations.
Protons are emitted within a cone of $5^{\circ}$ half-aperture angle because, as reported here\cite{2016MINEOreport} , above this angle the effective area to proton funneling falls rapidly and no protons exit towards the focal plane. Only the SS model is used in the present simulations.
\\
The correct assembly of the SPO full mass model is tested by computing the spatial, angular and energy distribution of protons at the optics exit and on the focal plane (Fig. \ref{fig:spatial}, \ref{fig:radial} and \ref{fig:energy}). The same exposure to the proton flux is simulated for each row. 
\\
The scattering efficiency, given by the ratio of exiting to entering protons in the the mirror modules, is 0.4\% for all particles and 0.03\% for the selection.
Simulations obtained with a ray-tracing code using the actual SPO design have shown that the conical approximation used for the building of the Geant4 Si plates transmits 20\% less photons\cite{private_mineo} . Assuming the same transmission reduction for protons, a 20\% uncertainty on the total proton flux on the focal plane can be accepted given the overall uncertainties of the input fluxes.
The transmission factor T is $(320\pm0.9)\times 10^{-7}$, with T defined as:
\begin{equation}
\rm T = \frac{N_{out}\times\Omega_{in}\times A_{SPO}}{N_{in}\times 4\pi\times A_{FP}}\;,
\end{equation}
and where N$_{\rm in}$ and N$_{\rm out}$ are the input and exiting protons, $\Omega_{\rm in}$ is the input cone solid angle, A$_{\rm SPO}$ and A$_{\rm FP}$ are the SPO surface, including the space between rows, and the 15 cm radius focal plane respectively.

\subsubsection{Spatial distribution}
Results are shown for all the protons exiting the optics and only the ones reaching the focal plane within 15 cm radius: while the diagonal side for the WFI is about 20 cm, the baffle entrance has a diameter of $\sim26$ cm for the WFI and $\sim22$ cm for the X-IFU. We will refer to this region as focal plane for simplicity, although the lateral size for the X-IFU is only few centimeters. 
A larger number of protons is scattered by the external rows, because of the smaller length of the P-H stack with respect to the inner rows (Fig. \ref{fig:spatial}, left panel), for the same focusing area. The focusing effect by the mirror is clear on the focal plane spatial distribution (Fig. \ref{fig:spatial}, right panel), where the central region is hit by about ten times the protons reaching a distance of 50 cm. The spatial and radial distribution on the 15 cm radius focal plane is better shown in Fig. \ref{fig:radial}. The radial distribution is plotted as surface density, with protons counted in rings, with length equal to the radial binning, and divided by the ring area. This density is reduced by 25\% from the center to the 15 cm radius edge. Most of protons have energies below 30 keV, where the optical filter would likely be able to stop them. If we consider the higher ranges, the decrease is well below 25\%. With respect to the XMM-Newton MOS observed radial distribution of scattered soft protons\cite{kun08}, where a factor 2 reduction is clearly observed from the center to the edge, ATHENA SPO configuration induces much lower vignetting.

\subsubsection{Energy and angular distribution}
Using the SS model for the description of the physics interaction, the energy loss percentage for protons exiting the ATHENA mirrors (Fig. \ref{fig:energy}, top-left panel) is mostly below 5-10\%, confirming the values obtained in the validation activity\cite{2017ExA....44..413F} .
\\
Fig. \ref{fig:energy} (top-right panel) shows the energy spectrum, given by each row and their sum, for protons reaching the 15 cm radius selection, normalized for the selected area. The external rows cause a flux 10 times higher than the inner ones, partially induced by the lower length of the P-H stacks and their larger area.  The total  residual flux is $\sim40$ prot. cm$^{-2}$ s$^{-1}$ keV$^{-1}$ at 10 keV, about $10^{4}$ times lower than the input flux at the same energy. If we assume a stopping power of $\sim20$ keV caused by the optical filters, as the case of XMM-Newton thin filter\cite{tie07}, the residual flux at 20 keV is $\sim3$ prot. cm$^{-2}$ s$^{-1}$ keV$^{-1}$.
\\
The polar angular distribution of each row is normalized for the input number of protons, in each row, and the bin width. If we consider all the protons exiting the SPO (Fig. \ref{fig:energy}, bottom-left panel), the polar angle peaks increasingly from $\sim1^{\circ}$ for the inner rows to $>4^{\circ}$ angles for the outer ones, but with a larger spread in the distribution, and with a higher scattering efficiency. If we only select the protons reaching the 15 cm radius focal plane, the scattering efficiency is around the same for all the rows and narrowly peaked in the $1^{\circ} - 7^{\circ}$ range.

%%%%%%%%%%%%%%%%%%%%%%%%%%%%%%%%%%%%%%%%%%%%%%%%%%%%%%%%%%%%%

\section{Summary}
The Geant4 mass model for the ATHENA SPO structure, the Wolter-I X-ray focusing telescope on board the next large ESA X-ray mission, is developed for the simulation of the interaction of charged particles with its Si pores. The unprecedented number of volumes composing the telescope requires the use of parameterised classes for each component, while the use of BoGEMMS configuration files allows to adapt the configuration to the evolving design without rebuilding a new mass model. We demonstrate that introducing the coating on the Si pore surface changes the total number of protons hitting the detectors of only few percents, so that omitting the coating volumes simplifies the mass model while not significantly impacting on the scattered proton distribution.
\\
The mass model verification tests, using both a single pore or the full SPO, allow to study the effect of different interaction models on the protons exiting the mirrors and to preliminary evaluate the residual proton flux on the focal plane. The Remizovich solution, in the approximation of no energy losses, and the Coulomb single scattering models result in a similar polar and azimuthal angular distribution, both with single and double reflections accounting for $\sim70\%$ of interactions. The first model, however, shows about a doubled scattering efficiency with respect to the SS model. 
\\
The soft proton flux, in the 10 - 500 keV energy range, modelled for the plasma sheet magnetotail region where highest rates are expected, is used as input for the evaluation of the total energy, angular and spatial distribution of protons scattered by the complete SPO mass model. While protons are clearly concentrated towards the central regions of the focal plane, a weak vignetting effect, lower that 25\% from the center to a 15 cm radius, is observed, contrary to the factor 2 reduction observed by the XMM-Newton MOS detectors. The residual proton flux with a 15 cm radius focal plane, using the SS model as interaction process, is $\sim40$ prot. cm$^{-2}$ s$^{-1}$ keV$^{-1}$ at 10 keV, about $10^{4}$ times lower than the input flux at the same energy. 

\section{Conclusions}
The SPO mass model has been delivered to ESA as product of the WP 7.2 of the AREMBES project, and it will made available to the users as part of the ATHENA mass model of the AREMBES Simulation Framework. Present simulations of soft proton scattering show that low energy protons in L2 can indeed be funnelled by the Si pores. Because of the large effective area of the telescope and the higher fluxes expected in L2, the long-term averaged residual flux on the focal plane, calculated at 20 keV to account for the presence of XMM EPIC thin filter, is of the order or higher than the most intense soft proton flares observed\cite{ken00} by XMM-Newton EPIC pn CCDs.
\\
The work presented here has been developed in collaboration and in constant verification with independent Geant4 simulations of soft proton scattering with ATHENA optics\cite{2017DUSANreport} and all results, including the residual flux on the focal plane, are consistent.

\acknowledgments     %>>>> equivalent to \section*{ACKNOWLEDGMENTS}       
 
The present work has been supported by the ESA AREMBES project, contract No 4000116655/16/NL/BW.

% References
\bibliography{fioretti_bib} % bibliography data in report.bib
\bibliographystyle{spiebib} % makes bibtex use spiebib.bst

\end{document}